\documentclass[a4paper]{article}

\usepackage{fullpage} 
\usepackage{pstricks}
\usepackage{multicol}
\usepackage{graphicx}
\usepackage{movie15}
\usepackage{hyperref}
\usepackage[version=3]{mhchem}
\usepackage{amsmath} 
\usepackage{amsfonts}
\usepackage{color}
\usepackage{soul}
\usepackage[margin=1.7cm,top=3cm,bottom=2.8cm,centering]{geometry}
\definecolor{rosepale}{rgb}{1.0, 0.7, 1.0}

\title{Conformational selection or induced fit? \\ New insights from old principles\footnote{Reference: Michel, D. 2016. Conformational selection or induced fit? New insights from old principles. Biochimie 128-129, 48-54}}

\author{Denis Michel \\
\\
      \begin{small} Universite de Rennes1-IRSET. Campus Sant\'e de Villejean. 35000 Rennes France. denis.michel@live.fr \end{small}
\\}

\date{} 

\begin{document}
\maketitle

\begin{multicols}{2}

\textbf{ABSTRACT}. A long standing debate in biochemistry is to determine whether the conformational changes observed during biomolecular interactions proceed through conformational selection (of preexisting isoforms) or induced fit (ligand-induced 3D reshaping). The latter mechanism had been invoked in certain circumstances, for example to explain the non-Michaelian activity of monomeric enzymes like glucokinase. But the relative importance of induced fit has been recently depreciated in favor of conformational selection, assumed to be always sufficient, predominant in general and in particular for glucokinase. The relative contributions of conformational selection and induced fit are reconsidered here in and out of equilibrium, in the light of earlier concepts such as the cyclic equilibrium rule and the turning wheel of Wyman, using single molecule state probability, one way fluxes and net fluxes. The conditions for a switch from conformational selection to induced fit at a given ligand concentration are explicitly determined. Out of equilibrium, the inspection of the enzyme states circuit shows that conformational selection alone would give a Michaelian reaction rate but not the established nonlinear behaviour of glucokinase. Moreover, when induced fit and conformational selection coexist and allow kinetic cooperativity, the net flux emerging in the linkage cycle necessarily corresponds to the induced fit path. 

\section{Introduction}

Folding and binding are intertwined processes involved in intra- and inter-molecular interactions. It is generally accepted that macromolecular interactions are rarely rigid (lock and key) and involve conformational changes. A persistent question in this context is whether these changes result from the selection of a subset of preexisting conformations (conformational selection: CS) or from a post-binding stereo-adjustment (induced fit: IF). \cite{Hammes2,Vogt2012,Changeux,Sullivan,Vogt2013,Vogt2014,Gianni}.
IF has been differentiated from CS through the mean times of equilibrium restoration following perturbation by changing substrate concentration. When assuming slow conformational changes and rapid ligand turnovers, as the ligand concentration increases, the relaxation time increases in IF and decreases in CS \cite{Hammes1}. Hence, relaxation times have logically been used as a diagnostic to appraise the relative contributions of CS and IF and led for instance to the conclusion that CS is predominant for glucokinase previously believed to obey IF \cite{Vogt2012}. The interpretation of relaxation times is however more delicate in absence of the poorly justified hypothesis of time scale separation between the transconformation and binding phenomena \cite{Vogt2014}. It could be usefully completed by complementary kinetic studies \cite{Meyer-Almes}. Alternatively, it has been proposed that the relative contribution of CS and IF can be evaluated by comparing their fluxes \cite{Hammes2}. It is this approach that is selected in this study. The CS and IF flows are compared in and out of equilibrium, in general and in the specific case of enzymatic reactions, which ideally illustrate the non-equilibrium situations in the cell. These driven reactions will be used as a thread of this article, to show that pure IF and CS are not topologically equivalent and to examine the relative contributions of CS and IF. The example of glucokinase will prove particularly instructive to evaluate the role of IF in kinetic cooperativity, which is an example of the essential roles ensured by CS and IF in the nonlinearity of biochemical phenomena.

\section{Roles of CS and IF in non-Michaelian behaviours}

CS and IF can stabilize specific folding states and underly nonlinear biochemical phenomena such as sigmoidal reactions. Sigmoidal or cooperative saturation curves are themselves essential for regulating biochemical systems and their possible multistability. Two main modes of sigmoidal saturation by a single ligand have been identified, which differ by the presence of a unique or several binding sites on the macromolecule. Multisite cooperativity, frequently encountered for regulatory enzymes, can proceed through either conformational selection or induced fit, illustrated by the MWC \cite{Monod} and KNF \cite{Koshland} models respectively. A general specificity of multisite cooperativity is to hold as well in equilibrium and nonequilibrium conditions, as long evidenced by the oxygenation of vertebrate hemoglobin that is not an enzyme. By contrast, single site cooperativity is more rarely observed and exists out of equilibrium only. These two modes of cooperativity have been pertinently called cooperativity through space (between the different sites at a given time point) and through time (between the successive states of the same site) \cite{Whitehead}. The latter model, of kinetic cooperativity, has been clearly described by Rabin \cite{Rabin}. Its principle can be visualized intuitively using the scheme of Fig.1.

\begin{center}
\includegraphics[width=7cm]{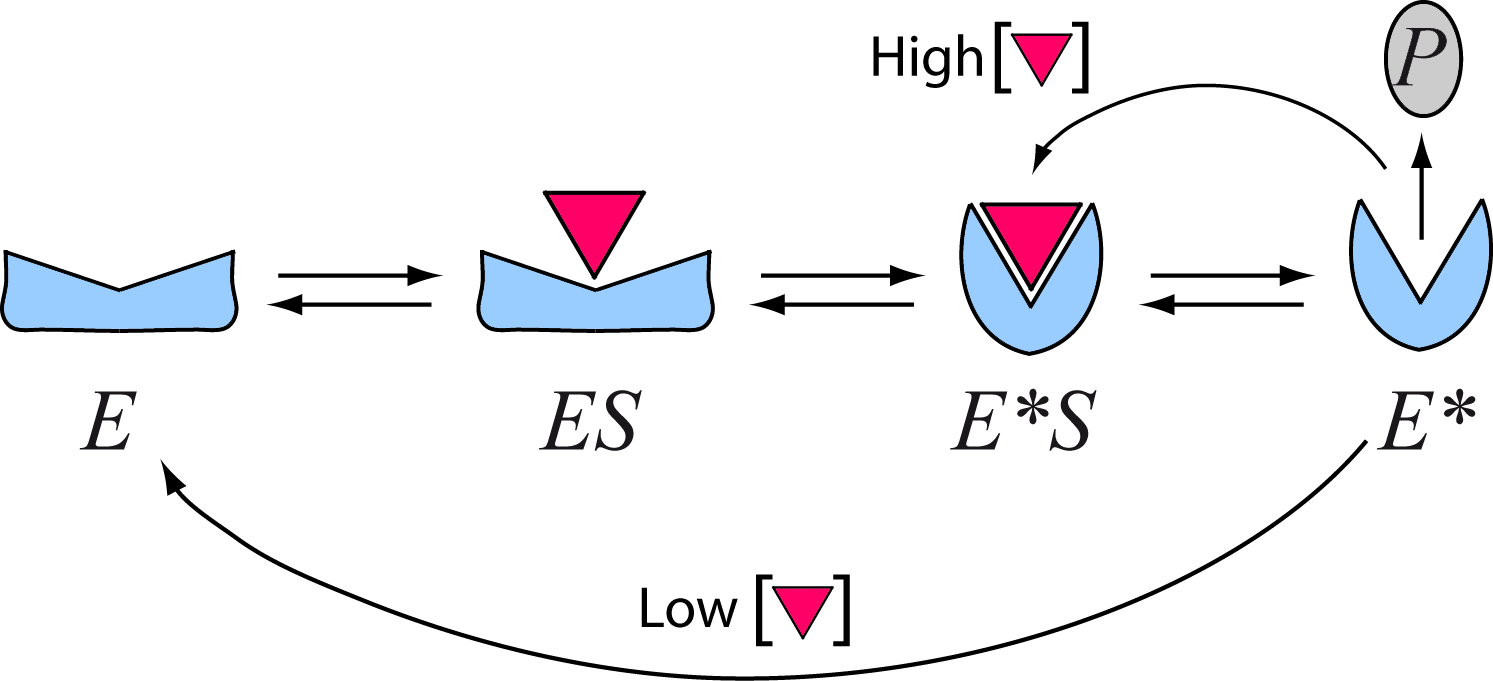} \\
\end{center}
\begin{small} \textbf{Figure 1.} The nonlinear dependence on the substrate of the enzyme activity can be explained by a competition between relaxation of $ E^{*} $ into $ E $ and the direct rebinding of the substrate to $ E^{*} $. At low substrate concentration (triangles), transconformation takes place, reinitiating the long cycle with its inefficient initial association, whereas at higher concentration of glucose, a molecule binds immediately to the enzyme. In this short cycle, the enzyme remains in its reactive form so that the global reaction rate is speeded up. \end{small}\\

In its basal conformation ($ E $), the enzyme binds the substrate with a low affinity, but once bound, the substrate induces the transconformation of the enzyme into a new form $ E^{*} $ that is "kinetically favourable but thermodynamically unfavourable" \cite{Rabin}, that is to say less stable than $ E $ in absence of substrate, but conveniently folded for catalyzing the reaction. After release of the reaction product (oval in Fig.1), a kinetic competition begins between two events: the relaxation of the enzyme to its basal conformation and the reattachment of a new substrate. For a fixed relaxation rate constant, the issue of this race depends on substrate concentration. \\
 The enzyme glucokinase/ATP-D-glucose 6-phosphotransferase EC 2.7.1.1-hexokinase D is a good candidate to follow this mechanism according to models and experimental evidences \cite{Cardenas1975,Storer1976,Storer1977,Cardenas1979}. This enzyme enriched in the liver and pancreas of vertebrates, displays a non-Michaelian reaction rate with respect to its substrate (glucose), that is not observed for its counterparts catalyzing the same reaction in other tissues. The physiological outcome of the non-Michaelian behaviour of glucokinace is remarkable as it greatly contributes to the regulation of blood glucose levels, as suggested by the critical role of glucokinase in preventing diabetes \cite{Vionnet}. All the hexokinases catalyse the transformation of glucose into glucose 6-phosphate (G6P), but the fate of G6P depends on the organs. While it is consumed in most tissues, the liver has the capacity to temporarily store it in the form of glycogen, but only if the concentration of glucose in the blood exceeds a certain threshold. This exquisite mechanism is based on an induced-fit mechanism. However, using the criterion of the relaxation times \cite{Vogt2012}, authors revised the importance of IF and suggested that CS is in fact predominant in general \cite{Changeux,Vogt2013} and in particular for glucokinase \cite{Vogt2012,Kim}, for which different models have been proposed, such as those compiled in \cite{Larion} and earlier ones \cite{Cardenas1979}. In the alternative approach proposed here, CS and IF are tested alone or in combination for their capacity to yield the well established non-Michaelian activity of glucokinase. While relaxation times are obtained from a transient return to pure IF and CS equilibrium \cite{Vogt2012,Hammes1}, enzymatic reaction rates are measured in driven non-equilibrium steady states. Besides, the predominance of CS over IF described in \cite{Changeux} concerns multimeric cooperativity in equilibrium (for hemoglobin) or under the quasi-equilibrium hypothesis (for enzymes like aspartate transcarbamylase \cite{Monod}), but the cooperativity of monomeric enzymes like glucokinase can not be obtained in this way \cite{Ricard}. In this study, the substrates will be considered as rigid small molecules. We will adopt the classical approximation, as old as the theory of Michaelis-Menten, that the substrates are much more numerous than the macromolecules, which allows using pseudo-first order constants of association.

\section{Pure CS and IF enzymatic reactions}

\subsection{Pure CS gives a traditional Michaelian velocity}
The reaction following pure CS is

\begin{center}
\ce{\textit{E}
<=>[\ce{\textit{b}}][\ce{\textit{r}}]
$\ce{\textit{E* + S}}$
<=>[\ce{\textit{a}}][\ce{\textit{d}}]
$\ce{\textit{E*S}}$
->[\ce{\textit{c}}]
$\ce{\textit{E*+P}}$
}
\end{center}

where $ b $ (time$ ^{-1} $), $ r $ (time$ ^{-1} $), $ a $ (concentration$ ^{-1} $ time$ ^{-1} $) and $ d $ (time$ ^{-1} $) are the rate constants of bending of $ E $ into $ E^{*} $, reversion to the basal conformation, association and dissociation respectively. As shown below, this scheme always gives a Michaelian reaction rate.

\subsubsection{Under the quasi-equilibrium approximation}

The relative concentration of the different forms of the enzyme are simply linked by equilibrium constants $ K_{b}=b/r $ and $ K_{a}=a/d $. The reaction rate is proportional to the fraction of enzyme in the form $ E^{*}S $ 

\begin{equation} v= c \ \dfrac{K_{a}K_{b}[S]}{1+K_{b}+K_{a}K_{b}[S]} \end{equation}

\subsubsection{In nonequilibrium steady state}

When $ c $ is not negligible, we obtain the familiar hyperbolic reaction rate of Briggs and Haldane

\begin{subequations} \label{E:gp}
\begin{equation} v=\dfrac{V_{M}[S]}{K_{M}+[S]} \end{equation} \label{E:gp1}
with a maximum velocity
\begin{equation} V_{M}=c \end{equation} \label{E:gp2}
and a Michaelis constant
\begin{equation} K_{M}=\dfrac{(b+r)(c+d)}{ab} \end{equation} \label{E:gp3}
\end{subequations}
CS has been shown predominant for glucokinase \cite{Vogt2012}, but we see here that it is clearly incapable alone to explain kinetic cooperativity. Let us now examine the pure induced fit mechanism. 

\subsection{The pure induced fit mechanism is necessarily cyclical}

At first glance, the scheme

\begin{center}
\ce{\textit{E+S}
<=>[\ce{\textit{a}}][\ce{\textit{d}}]
$\ce{\textit{ES}}$
<=>[\ce{\textit{b}}][\ce{\textit{r}}]
$\ce{\textit{E*S}}$
->[\ce{\textit{c}}]
$\ce{\textit{E* + P}}$}
\end{center}

seems symmetrical to the CS scheme by permuting the conformational transition and the binding reaction; but in fact the above scheme  is clearly incomplete because the enzyme should recover its basal conformation to bind a new substrate in case of pure IF. An additional backward transition ($ r^{*} $) is thus necessary to reconvert the final form $ E^{*} $ into the initial form $ E $, thereby giving a cycle. Like the CS scheme, this minimal IF cycle is merely Michaelian, with new expressions for $ V_{M} $ and $ K_{M} $.

\begin{subequations} \label{E:gp}
\begin{equation} V_{M}=\dfrac{cbr^{*}}{bc+(b+c+r)r^{*}} \end{equation} \label{E:gp1}
and
\begin{equation} K_{M}=\dfrac{(bc+cd+dr)r^{*}}{a(bc+(b+c+r)r^{*})} \end{equation} \label{E:gp2}
\end{subequations}

In this pure IF scheme, the isoform $ E^{*} $ does not accommodate directly a new substrate molecule, but there is no obvious reason that a substrate can not bind to $ E^{*} $ before it relaxes. This would give the full CS/IF cycle described below.

\section{The IF/CS dual cycle}

\subsection{The IF/CS dual cycle at equilibrium}

The complete cycle of enzyme state transitions is represented in Fig.2.

\begin{center}
\includegraphics[width=8cm]{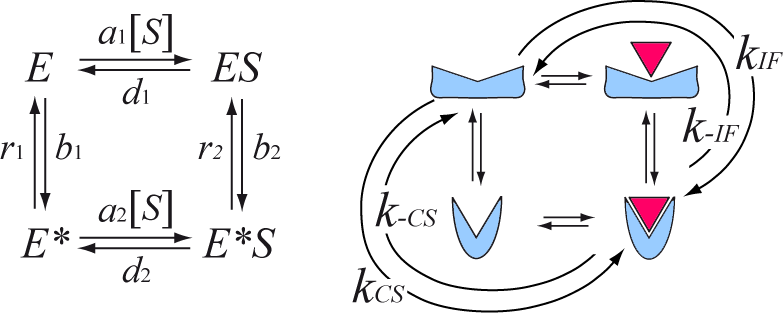} \\
\end{center}
\begin{small} \textbf{Figure 2.}  First-order binding scheme coupling CS and IF. $ k_{CS} $, $ k_{-CS} $, $ k_{IF} $ and $ k_{-IF} $ are the global rates of CS, reverse CS, IF and reverse IF respectively. \end{small}\\

This cycle is more realistic than pure CS or IF because even if certain enzyme states have a very low probability, they are nonetheless not forbidden.

\subsubsection{The CS-IF switch}
The relative importance of CS vs IF in equilibrium has been evaluated through different ways, including experimental relaxation times \cite{Vogt2012} and one-way fluxes \cite{Hammes2}. However, this predominance may be not absolute if it can pass from one mechanism to another depending on the conditions. Precisely, authors proposed that a higher ligand concentration could favor IF \cite{Hammes2,Daniels}, as supported by simulations, provided the conformational transition is slow enough \cite{Greives}. The dependence on the substrate of the relative contribution of CS and IF can be explicitly predicted by single molecule probability. This method is based on the comparative probability that a single molecule $ E $ reaches the state $ E^{*}S $ via either $ E^{*} $ (CS, with a global rate $ k_{CS} $) or via $ ES $ (IF, with a global rate $ k_{IF} $) (Fig.2). $ k_{CS} $ and $ k_{IF} $ are the reciprocal of the mean times of first arrival to $ E^{*}S $ and can be understood as "conditional rates" as follows. In the example of the CS path, the global rate is the rate that $ E $ commits first to $ E^{*} $ and that once at this state, it continues forward until $ E^{*}S $ instead of reverting back to $ E $. The probability of this event is $ a_{2}[S]/(r_{1}+a_{2}[S]) $ \cite{Malygin}, which gives

\begin{subequations} \label{E:gp}
\begin{equation} k_{CS} = b_{1} \dfrac{a_{2}[S]}{r_{1}+a_{2}[S]} \end{equation} \label{E:gp1}
and in the same manner
\begin{equation} k_{IF} = a_{1}[S] \dfrac{b_{2}}{d_{1}+b_{2}} \end{equation} \label{E:gp2}
\end{subequations}

$ k_{CS} $ and $ k_{IF} $ are hyperbolic and linear functions of $ [S] $ respectively. Obviously these functions can intersect if the initial slope of the hyperbola ($ b_{1} a_{2}/r_{1} $) is higher than the slope of $ k_{IF} $, which is achieved when the relaxation rate constant is low enough such that

\begin{equation} \dfrac{r_{1}}{b_{1}} < \dfrac{a_{2}}{a_{1}} \left (1+\dfrac{d_{1}}{b_{2}} \right ) \end{equation}

In this case, there is a certain value of $ [S] $ below which $ k_{CS}> k_{IF} $ and over which $ k_{CS} < k_{IF} $. The critical substrate concentration is
\begin{equation} [S]  = \dfrac{b_{1}}{a_{1}}\left (1+\dfrac{d_{1}}{b_{2}}  \right )-\frac{r_{1}}{a_{2}} \end{equation}

The above global rates of CS and IF for a single molecule $ E $  are always valid, but it is also interesting to consider the number of molecules to which they apply. The product of single molecule rates by the concentration of the molecules involved, is called a flux. Expectedly, the one-way flux approach of \cite{Hammes2} gives the same result. Indeed, the one-way fluxes passing through the CS and IF specific branches of Fig.2 are

\begin{equation} J_{CS} = \left (\frac{1}{b_{1}[E]}+\frac{1}{a_{2}[E^{*}][S]} \right )^{-1} \end{equation}
and
\begin{equation} J_{IF}  = \left (\frac{1}{a_{1}[E][S]}+\frac{1}{b_{2}[ES]} \right )^{-1} \end{equation}
When replacing $ [E^{*}] $ by $ b_{1}[E]/r_{1} $ and $ [ES] $ by $ a_{1}[E][S]/r_{1} $, solving $ J_{CS} = J_{IF} $ naturally gives the same critical value of $ [S] $ as above.\\

When defining the global rates of reverse CS ($ k_{-CS} $) and reverse IF ($ k_{-IF} $) (Fig.2), the identity of forward and backward CS and IF fluxes in equilibrium can be readily shown. On the one hand, the global path rates are
\begin{itemize}
\item $ k_{CS}=b_{1}a_{2}[S]/(r_{1}+a_{2}[S]) $
\item $ k_{-CS}=d_{2}r_{1}/(r_{1}+a_{2}[S]) $
\item $ k_{IF}=b_{2}a_{1}[S]/(d_{1}+b_{2}) $
\item $ k_{-IF}= d_{1}r_{2}/(d_{1}+b_{2}) $
\end{itemize}
and on the other hand, the relative concentrations of the different forms of the enzyme are related through equilibrium constants in equilibrium, as for example $ [E^{*}S]=\dfrac{b_{1}}{r_{1}}\dfrac{a_{2} }{d_{2}} [E][S] $.
Coupling these relationships to the global rates implies

$$ k_{-CS}[E^{*}S]=\dfrac{a_{2}b_{1}[E][S]}{r_{1}+a_{2}[S]}=k_{CS}[E]  $$

Hence, we have

\begin{subequations} \label{E:gp}
\begin{equation} k_{CS}[E]= k_{-CS}[E^{*}S] \end{equation} \label{E:gp1}
and
\begin{equation} k_{IF}[E]= k_{-IF}[E^{*}S] \end{equation} \label{E:gp2}
\end{subequations}

The possible predominance of a path, either CS or IF, is the same for complex formation and dismantlement, owing to microscopic reversibility.

\subsubsection{The CS-IF detailed balance}

The generalized reversibility of forward and backward fluxes at equilibrium is known as the detailed balance and holds for any transition, as minor as it can be, in a network of any size \cite{Lewis}. As a consequence for the cyclic network of Fig.2, the 8 rate constants are mutually constrained by the detailed balance rule (Appendix A) reading

\begin{subequations} \label{E:gp}
\begin{equation} a_{1}b_{2}d_{2}r_{1}=a_{2}b_{1}d_{1}r_{2} \end{equation} \label{E:gp1}
or differently written with the equilibrium constants
\begin{equation} \dfrac{K_{a1}}{K_{a2}}=\dfrac{K_{b1}}{K_{b2}} \end{equation} \label{E:gp2}
\end{subequations}

Studies aimed at determining the relative importance of CS and IF, are plausible only if the values of the constants satisfy Eq.(10). This criterion is for example verified for the rate constants given for Flavodoxin in \cite{Hammes2}, but strangely not in all the models developed for glucokinase. At equilibrium, the clockwise (IF) and counterclockwise (CS) fluxes of Fig.2 cancel each other and the saturation curve is a hyperbolic function of the substrate:

\begin{subequations} \label{E:gp}
\begin{equation} Y=\dfrac{[S]}{K_{d}^{app}+[S]} \end{equation} \label{E:gp1}
with
\begin{equation} K_{d}^{app}=\dfrac{1+K_{b1}}{(1+K_{b2})K_{a1}} \end{equation} \label{E:gp2}
\end{subequations}
and the enzymatic reaction rate is once again Michaelian. The reaction rate is proportional to the probability for a single enzyme to be in state $ E^{*}S $ 

\begin{equation} v=c \ P(E^{*}S) \end{equation}

with

\begin{equation} P(E^{*}S)=\dfrac{[E^{*}S]}{[E]+[E^{*}]+[ES]+[E^{*}S]} \end{equation}
where each concentration can be expressed as a function of $ [E] $ and equilibrium constants as shown above, so that the fraction can finally be simplified by eliminating $ [E] $ and gives the Briggs-Haldane velocity $ v=V_{M}[S]/(K_{d}^{app}+[S]) $ with $ K_{d}^{app} $ defined above and $ V_{M}=c \ K_{b2}/ (1+K_{b2}) $. We verify that the nonlinear mechanism of conformational memory does not exist in equilibrium \cite{Ricard}.

\subsection{The mixed IF/CS cycle in non-equilibrium steady state}

Equilibrium can be broken if the enzymatic rate constant $ c $ is not negligible compared to the other rate constants. In this case, a net circulation appears in the cycle, clockwise with respect to Fig.3 and whose value is

\begin{subequations} \label{E:gp}
\begin{equation}  J_{NF}=\dfrac{cr_{1}b_{2}a_{1}[S]}{\mathcal{D}} \end{equation} \label{E:gp1}
with
\begin{equation}\begin{split} \mathcal{D}&=(r_{1}+b_{1})(d_{1}r_{2}+(d_{1}+b_{2})(d_{2}+c))\\ &+(r_{1}(r_{2}+b_{2})+(d_{2}+c)(r_{1}+b_{2})) \ a_{1}[S] \\&+(b_{1}(d_{1}+b_{2})+r_{2}(d_{1}+b_{1})) \ a_{2} [S]\\&+(r_{2}+b_{2}) \ a_{1}a_{2}[S]^{2} \end{split}\end{equation} \label{E:gp2}
\end{subequations}

\begin{center}
\includegraphics[width=8cm]{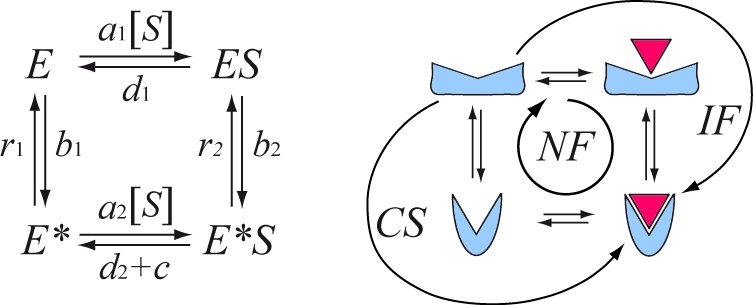} \\
\end{center}
\begin{small} \textbf{Figure 3.}  First-order scheme of enzyme state recycling, mixing CS, IF and product release (rate $ c $). NF: net flux arising out of equilibrium (the "turning wheel" of Wyman \cite{Wyman}) \end{small}\\

$ J_{NF} $ is maximal at the substrate concentration cancelling the derivative of this function (Appendix B). This flux holds for all the transitions, in particular for the IF transition ($ b_{2}[ES]-r_{2}[E^{*}S] $), which is strictly positive for nonzero $ c $. In other words, the flux $ E\rightarrow ES \rightarrow E^{*}S $ is stronger than $ E^{*}S \rightarrow ES \rightarrow E $ whereas $ E\rightarrow E^{*} \rightarrow E^{*}S $ is lower than $ E^{*}S \rightarrow E^{*} \rightarrow E $. With this net flow around the cycle, non-Michaelian behaviours arise. For instance, the fraction of occupation of the enzyme by the substrate is no longer a hyperbola but becomes a nonlinear function of the substrate (Appendix C). The enzymatic velocity takes the general form 

\begin{equation} v=c \ \dfrac{A[S]+B[S]^{2}}{C+(A+D)[S]+B[S]^{2}} \end{equation}

where 

\begin{itemize}
\item $ A=a_{1}b_{2}r_{1}+a_{2}b_{1}(d_{1}+b_{2}) $
\item $ B=a_{1}a_{2}b_{2} $
\item $ C=(b_{1}+r_{1})(d_{1}r_{2}+(d_{1}+b_{2})(d_{2}+c)) $
\item $ D= a_{1}(r_{1}r_{2}+(b_{2}+r_{1})(d_{2}+c))+a_{2}r_{2}(b_{1}+d_{1}) $
\end{itemize}

This formula can give the expected sigmoidal rate of glucokinase for a range of parameters. To quantify this sigmoidicity, the extent of cooperativity is classically evaluated through the Hill coefficient $ n_{H} $. It is the extreme slope (minimal or maximal) of the so-called Hill plot drawn in Logit coordinates. Starting from
\begin{subequations} \label{E:gp}
\begin{equation} \dfrac{v}{V_{M}}=\dfrac{A[S]+B[S]^{2}}{C+(A+D)[S]+B[S]^{2}} \end{equation} \label{E:gp1}
\begin{equation}\dfrac{v}{V_{M}-v}= \dfrac{A[S]+B[S]^{2}}{C+D[S]}, \end{equation} \label{E:gp2}
the Hill function $ H = \ln (v/(V_{M}-v))$ can be examined directly in Logit coordinates. If writing $ x=\ln [S] $, its first and second derivatives are
\begin{equation}  H'(x)=1-\dfrac{A}{A+B\textup{\large{e}}^{x}} +\dfrac{C}{C+D\textup{\large{e}}^{x}}\end{equation} \label{E:gp3}
\begin{equation} \begin{split} H''(x)=& \dfrac{A}{A+B\textup{\large{e}}^{x}}-\dfrac{A^{2}}{(A+B\textup{\large{e}}^{x})^{2}}\\& -\dfrac{C}{C+D\textup{\large{e}}^{x}}+\dfrac{C^{2}}{(C+D\textup{\large{e}}^{x})^{2}}   \end{split} \end{equation} \label{E:gp4}
\end{subequations}
The substrate concentration giving the extreme slope is obtained by solving $ H''(x)=0 $

\begin{equation} [S]= \sqrt{\dfrac{AC}{BD}} \end{equation} 

giving, when reintroduced into $ H' $, a Hill coefficient of

\begin{equation} n_{H}= \dfrac{2}{1+\sqrt{\dfrac{AD}{BC}}} \end{equation} 

In the present case, $ n_{H} $ cannot exceed 2. It is higher than 1 (sigmoidal dependance on substrate) for values of the constants such that $ BC>AD $. Positive cooperativity (Sigmoid) is precisely obtained for $ n_{H} > 1 $, that is to say when $ BC > AD $. Interestingly for glucokinase, a Hill coefficient of about $ n_{H}=1.5 $ has been measured, but at low ATP concentrations (when $ c \approx 0 $), $ n_{H}=1 $  \cite{Storer1976,Cardenas1979}, confirming the Michaelian activity in absence of the net cyclical flux, that is the primary condition for obtaining kinetic cooperativity through conformational memory. To conceive the importance of its cyclical nature, note for comparison that no kinetic cooperativity is possible in the linear scheme of pure CS described previously, in which $ E^{*} $ had also the dual possibility to either relax or catch a substrate molecule. The net cyclical flux should also be oriented in the direction of the IF path. As a matter of fact, the general form of velocity classically admitted for glucokinase is not obtained if we artificially force the net flux to be counterclockwise (Appendix D). Hence, CS is clearly not sufficient in the present context and the positive net IF flux is necessary for kinetic cooperativity to appear, unless the accepted conclusion of conformational cooperativity is wrong for glucokinase and that its behaviour results in fact from a completely different mechanism not based on conformational changes \cite{Pettersson}. This remaining possibility is described below.

\section{Monomeric sigmoidicity without IF and CS}

The above considerations clearly show that CS is insufficient to explain the sigmoidal dependence on glucose of glucokinase through the model of kinetic cooperativity. Hence, it is necessary to examine if alternative models can give this result while allowing dominant CS. One such model exists, this is the random binding of different substrates on an enzyme catalyzing bimolecular reactions. It could in principle apply to glucokinase which has two substrates: glucose and ATP. The fixation of two substrates represented in Fig.4 is theoretically sufficient to give rise to non-Michaelian rates \cite{Dalziel} and, in particular, to sigmoidal rates \cite{Dalziel,Ferdinand}. 

\begin{center}
\includegraphics[width=5cm]{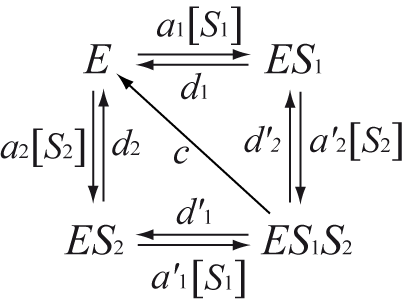} \\
\end{center}
\begin{small} \textbf{Figure 4.} Random building of a ternary complex containing the enzyme and two different substrate molecules.  \end{small}\\

Though simple, this system is subtle enough: (i) The completely hierarchical binding of the substrates, for example $ S_{1} $ always before $ S_{2} $, gives a Michaelian reaction rate. (ii) A completely random filling also leads to a Michaelian behaviour. (iii) But interestingly, the incomplete dominance or preference for one substrate, gives non-Michaelian curves: either sigmoidal or bell-shaped with a maximum, depending on the relative values of the constants. The conditions giving these results, the corresponding inflexion points and the apparent Hill coefficients, are detailed in \cite{Ferdinand}. For this ternary complex mechanism, conformational changes are possible but dispensable. Similar results are indeed obtained with rigid binding, for example in the case of asymmetrically charged ligands \cite{Ferdinand}. \\
This origin of cooperativity has been ruled out for glucokinase \cite{Cornish-Bowden}, since glucose binds before ATP to glucokinase and the binding of ATP remains Michaelian for all the doses of glucose. Only glucose causes a significant reshaping of the enzyme, in presence or absence of ATP and without consequence for its affinity for ATP. Large concentrations of ATP which would alter the binding hierarchy of the substrates, do not affect the cooperativity with glucose \cite{Cardenas1979}. Finally, the mutations (Y214A and N166R) which decrease cooperativity also lead to an increase of the affinity of glucokinase for glucose \cite{Moukil}. Together, these observations strongly support the model of kinetic cooperativity for glucokinase. 

\section{Conclusions}

As proposed in Hammes (2009), the relative contributions of CS and IF should be evaluated through fluxes, and the present study provides clear conclusions on this point, in and out of equilibrium. The ligand concentration point of switching between CS and IF is determined. Beside the symmetrical fluxes of equilibrium, the long established cycle encompassing both CS and IF is completed here with a net flux oriented along the IF pathway and responsible for a nonlinear substrate dependence. CS is definitely insufficient to yield the kinetic cooperativity of glucokinase given that (i) Pure CS and IF modes of binding are unable to explain the sigmoidal dependence on glucose of glucokinase, suggesting the existence of a mixed cycle combining CS and IF. (ii) In the dual cycle under the rapid pre-equilibrium assumption, the relative importance of CS and IF is joined by the detailed balance to the relative affinities of the two isoforms for the substrate. In this condition, kinetic cooperativity is impossible. (iii) In non-equilibrium steady state, a net cyclical flow appears, corresponding to the IF path, opposite to the CS path and allowing kinetic cooperativity. Hence, the claim that CS is always sufficient seems to not apply to kinetic cooperativity, which primarily relies upon IF. This crucial role of IF does not exclude the existence of pre-equilibria between many isoforms. Dynamic disorder is a widespread property of protein folding which may interfere with binding and the present report is not aimed at minimizing the importance of CS. The categories of enzyme states defined here ($ E $, $ E^{*} $, $ ES $ and $ E^{*}S $) can be viewed as representative of many sub-isoforms without altering the results. For example, fluctuating Michaelian enzymes still display globally Michaelian behaviours as long as the topology of the network and the associated net fluxes are maintained \cite{Min}. In the enzymatic reaction studied here involving the joined CS/IF scheme, the driver transition breaking equilibrium is the rate constant $ c $. It has been considered completely irreversible for simplicity, but a reversible transition unbalanced enough to break equilibrium would have been sufficient. The simplified one-way arrow commonly used in enzymology without corruption of the results, is not elementary since it covers at least two more elementary transitions: the transformation of substrate(s) into product(s) and the subsequent release of the product(s). The apparent irreversibility of enzymatic reactions in the usual assay conditions is not due to the fact that enzymes work in a one-way manner with a null rate constant for the reaction reverse to the catalysis, as often assumed, but simply reflects the negligible steady state product concentration. Indeed, a reaction is evolutionary selected when its product is useful for subsequent reactions or syntheses. As a consequence, the stationary concentration of the product is very low in the cell, thus preventing its reuptake by the enzyme. This pumping must of course be compensated by a permanent replenishment in substrate reflecting the open nature of cellular systems. This situation is mimicked in vitro when measuring initial velocities, when the concentration of products in the mixture is still negligible. In the context of the present report, the micro-irreversibility of $ c $ has two essential virtues: (i) it breaks the cyclic detailed balance rule and (ii) it directs the net flow. The topology of the steady state scheme of Fig.3 and its net IF flux, illustrate well the organizational potential of non-equilibrium mechanisms \cite{Michel2011} and the sentence of Prigogine "Nature begins to see out of equilibrium". In the present example, a single molecule of glucokinase, without the need for regulating its synthesis or of other biochemical circuits, can act as (i) a sensor of glucose concentration (a role also suggested in the pancreas), (ii) an enzyme with a conditional activity and (iii) a regulator of glucose concentration in the blood, working in a real time manner but not by corrective feedback. These properties may explain the striking evolutionary conservation of the Hill coefficient of glucokinase \cite{Cardenas1979} and the importance of this enzyme in the regulation of glycemia \cite{Vionnet}. For monomeric enzymes, this refined behaviour necessitates a non-equilibrium mechanism, that is to say a net flux reflecting a difference of free energy between the different forms of the enzyme. This difference is itself directly related to the asymmetry between forward and backward fluxes \cite{Beard}.

\begin{equation} G(E^{*}S)-G(E) = -RT \ln(J_{IF}/J_{-IF}) \end{equation}
\noindent
(Appendix E) which is of course zero in equilibrium and negative in driven steady state. It may seem surprising that similar behaviours are obtained in equilibrium for multimeric enzymes \cite{Michel2011}. Multimerization can be viewed as an optimized negentropic recipe in which the necessity for a driven system has been replaced by an upgraded information about protein structure \cite{Michel2013}.

\end{multicols}
\newpage
\begin{center}
\Huge{Appendices}
\end{center}
\appendix
\setcounter{equation}{0}  % reset counter 
\numberwithin{equation}{section}

\section{The cyclic equilibrium rule}
When the enzyme-substrate mixture is in equilibrium, the detailed balance rule or generalized micro-reversibility for cycles (Eq.(10)), also known as Wegscheider's relation, directly ensues from the product of the rate constant ratios which are related to the energies of each species, such as $ b_{1}/r_{1}=\textup{\large{e}}^{[\varepsilon(E)-\varepsilon(E^{*})]} $. Hence, the circular product of these ratios \cite{Michel2011} is

$$ \dfrac{b_{1}}{r_{1}}\dfrac{a_{2}}{d_{2}}\dfrac{r_{2}}{b_{2}}\dfrac{d_{1}}{a_{1}}=\textup{\large{e}}^{[\varepsilon(E)-\varepsilon(E^{*})]+[\varepsilon(E^{*})-\varepsilon(E^{*}S)]+[\varepsilon(E^{*}S)-\varepsilon(ES)]+[\varepsilon(ES)-\varepsilon(E)]}=\textup{\large{e}}^{0}=1   $$

An equivalent and even simpler way is to express the equilibrium constants as concentration ratios. Based on the familiar relationship

\begin{equation} \Delta G = -RT \ln K_{a1}+RT \ln \frac{[ES]}{[E][S]} \end{equation}

it is clear that at equilibrium, when $ \Delta G=0 $, $$ K_{a1} =\frac{[ES]}{[E][S]} $$
in the same manner,
$$ K_{b2} =\frac{[E^{*}S]}{[ES]}, K_{a2} =\frac{[E^{*}S]}{[E^{*}][S]}, K_{b1} =\frac{[E^{*}]}{[E]}$$

so that after simplification, the product of the constants expressed in this form gives

$$ K_{a1} K_{b2}= K_{a2} K_{b1} = \frac{[E^{*}S]}{[E][S]} $$

\section{Substrate concentration giving the maximal net flux}

$ J_{NF} $ has the general form

\begin{subequations} \label{E:gp}
\begin{equation} J_{NF}=\dfrac{\alpha [S]}{\beta + \gamma [S] + \delta [S]^{2}}  \end{equation} \label{E:gp1}
where the constants are given in Eq.(14). It is maximal when
\begin{equation} [S]=\sqrt{\beta /\delta} \end{equation} \label{E:gp2}
giving the flux 
\begin{equation} J_{NF}=\dfrac{\alpha }{\gamma + 2 \sqrt{\beta \delta}}  \end{equation} \label{E:gp3}
\end{subequations}

\section{Enzyme occupancy by the substrate in steady state}
\begin{equation} Y=P(ES)+P(E^{*}S)=\dfrac{A[S]+B[S]^{2}}{C+(A+D)[S]+B[S]^{2}}  \end{equation}
where $ A $, $ B $, $ C $, $ D $ are the following constants: 
\begin{itemize}
\item $ A=a_{1}r_{1}(b_{2}+r_{2}+d_{2}+c)+a_{2}b_{1}(b_{2}+r_{2}+d_{1}) $
\item $ B=a_{1}a_{2}(b_{2}+r_{2}) $
\item $ C=(b_{1}+r_{1})(d_{1}r_{2}+(b_{2}+d_{1})(d_{2}+c)) $
\item $ D= a_{1}b_{2}(d_{2}+c)+a_{2}d_{1}r_{2}$
\end{itemize}

\section{Comparative results of IF and CS net fluxes on enzymatic velocity}

The net flux arising in the mechanism of kinetic cooperativity is clockwise with respect to Fig.3. Let us push the asymmetry of the flow to its maximum. To this end, $ E^{*} $ is supposed to be a high energy species which rarely appears spontaneously from $ E $ ($ b_{1}=0 $), but is induced by the substrate. In turn, $ E^{*}S $ rarely reverses to $ ES $ owing to the stabilizing effect of the substrate ($ r_{2}=0 $). In the same way, the tightly bound substrate rarely leaves the complex $ E^{*}S $ unless it is converted into a product of different shape/charge ($ d_{2}=0 $). The absence of spontaneous substrate dissociation is logical when the binding site closes after wrapping around the substrate. In these simplified but acceptable conditions, the reaction rate reads

\begin{equation} v=\dfrac{ca_{1}b_{2}(r_{1}[S]+a_{2}[S]^{2})}{cr_{1}(d_{1}+b_{2})+(b_{2}c+b_{2}r_{1}+cr_{1})a_{1}[S]+b_{2}a_{1}a_{2}[S]^{2}} \end{equation}

which can be fitted to the observed sigmoidal glucose concentration-dependent rate of glucokinase. For comparison, if we artificially force the net flux to be clockwise, in the orientation of the CS path, by cancelling $ r_{1} $ and $ b_{2} $, we would obtain a bell-shaped velocity curve vanishing for high substrate concentration, in contradiction with all the observations.

\begin{equation} v=\dfrac{cd_{1}b_{1}a_{2}[S]}{d_{1}b_{1}(r_{2}+c)+(d_{1}r_{2}+b_{1}r_{2}+d_{1}b_{1})a_{2}[S]+r_{2}a_{1}a_{2}[S]^{2}} \end{equation}

\section{The flux-energy relationship}

Eq.(A.1) can be rearranged as

\begin{equation} \Delta G = -RT \ln \dfrac{a_{1}}{d_{1}} \dfrac{[E][S]}{[ES]} \end{equation}
that is simply
\begin{equation} \Delta G = -RT \ln(J_{+}/J_{-}) \end{equation}

\end{document}